\documentclass[12pt,preprint]{emulateapj}

\begin{document}

\shorttitle{Thermal X-rays from Millisecond Pulsars}
\shortauthors{Bogdanov et al.}

\title{Thermal X-rays from Millisecond Pulsars: \\ 
Constraining the Fundamental Properties of Neutron Stars}

\author{Slavko Bogdanov, Jonathan E. Grindlay, and George B. Rybicki} 

\affil{Harvard-Smithsonian Center for Astrophysics, 60 Garden Street,
Cambridge, MA 02138; \\ sbogdanov@cfa.harvard.edu, josh@cfa.harvard.edu}

\begin{abstract}

  We model the X-ray properties of millisecond pulsars (MSPs) by
  considering hot spot emission from a weakly magnetized neutron star
  (NS) covered by a hydrogen atmosphere.  We investigate the
  limitations of using the thermal X-ray pulse profiles of MSPs to
  constrain the mass-to-radius ($M/R$) ratio of the NS. The accuracy
  is strongly dependent on the viewing angle and magnetic inclination
  but is ultimately limited only by photon statistics. We demonstrate
  that valuable information regarding NSs can be extracted even from
  data of fairly limited photon statistics through modeling of
  archival observations of the nearby isolated PSRs J0030+0451 and
  J2124--3358. The X-ray emission from these pulsars is consistent
  with the presence of an atmosphere and a dipolar field
  configuration.  For both MSPs, the favorable geometry allows us to
  place limits on the allowed $M/R$ of NSs. Assuming 1.4 M$_{\odot}$,
  the stellar radius is constrained to be $R > 9.4$ km and $R > 7.8$
  km (68\% confidence) for PSRs J0030+0451 and J2124--3358,
  respectively. We explore the prospects of using future observatories
  such as \textit{Constellation-X} and \textit{XEUS} to conduct X-ray
  timing searches for MSPs not detectable at radio wavelengths due to
  unfavorable viewing geometry. We are also able to place strong
  constraints on the magnetic field evolution model proposed by
  Ruderman. The pulse profiles indicate that the magnetic field of an
  MSP does not have a tendency to align itself with the spin axis nor
  migrate towards one of the spin poles during the low-mass X-ray
  binary phase.

\end{abstract}


\section{Introduction}

Recent X-ray studies have revealed that a number of known
rotation-powered millisecond pulsars (MSPs) exhibit predominantly
thermal soft X-ray emission
\citep{Grind02,Zavlin06,Bog06a,Zavlin07}. The infered effective
emission radii $R_{\rm eff}$ indicate that this radiation is localized
in regions on the neutron star (NS) surface much smaller than the
expected stellar radius ($R_{\rm eff}\ll R$) but comparable to the
classical radius of the pulsar magnetic polar cap $R_{pc}=(2\pi
R/cP)^{1/2}R$. This finding is in agreement with theoretical models of
pulsars, which indicate that the conditions in the magnetosphere of a
typical MSP favor heating of the polar caps to $\sim$$10^6$ K by a
return flow of energetic particles along the open magnetic field lines
\citep[see, e.g.,][for details]{Hard02,Zhang03}. As this heat is
restricted to a small fraction of the NS, study of the X-ray spectra
and pulse profiles of MSPs can reveal important information about the
star such as the radiative properties of the NS surface, magnetic
field geometry, and NS compactness ($R/R_S$, where
$R_S=2GM/c^2$). This approach, originally proposed by \citet{Pavlov97}
in the context of radio MSPs, can serve as a valuable probe of key NS
properties that are inaccessible by other observational means
(e.g.~radio pulse timing). As shown by \citet{Pavlov97},
\citet{Zavlin98}, and \citet{Bog07}, a model of polar cap thermal
emission from an optically-thick hydrogen (H) atmosphere provides an
excellent description of the X-ray pulse profiles of PSR J0437--4715,
the nearest known MSP. On the other hand, a blackbody model is
inconsistent with the X-ray timing data and can be definitively ruled
out. Furthermore, there is compelling evidence for a magnetic dipole
axis offset from the NS center \citep{Bog07}. Finally, the compactness
of PSR J0437--4715 is constrained to be $R/R_S>1.6$ (99.9\%
confidence).  Thus, modeling of X-ray data of MSPs appears to be a
very promising approach towards answering long-standing questions
regarding the fundamental properties of MSPs and NSs, in general.

The present paper represents an extension of the work described in
\citet{Bog07}. Herein we explore the detailed properties of the MSP
X-ray emission model with particular emphasis on its use for
constraints on the NS equation of state (EOS). The work is organized
as follows. In \S2 we examine the properties of our model; in \S3 we
discuss an application of our model to archival X-ray observations of
PSRs J0030+0451 and J2124--3358. In \S4 we present a discussion and
end with conclusions in \S5.

%
%
\begin{figure*}[t!]
\includegraphics[width=0.98\textwidth]{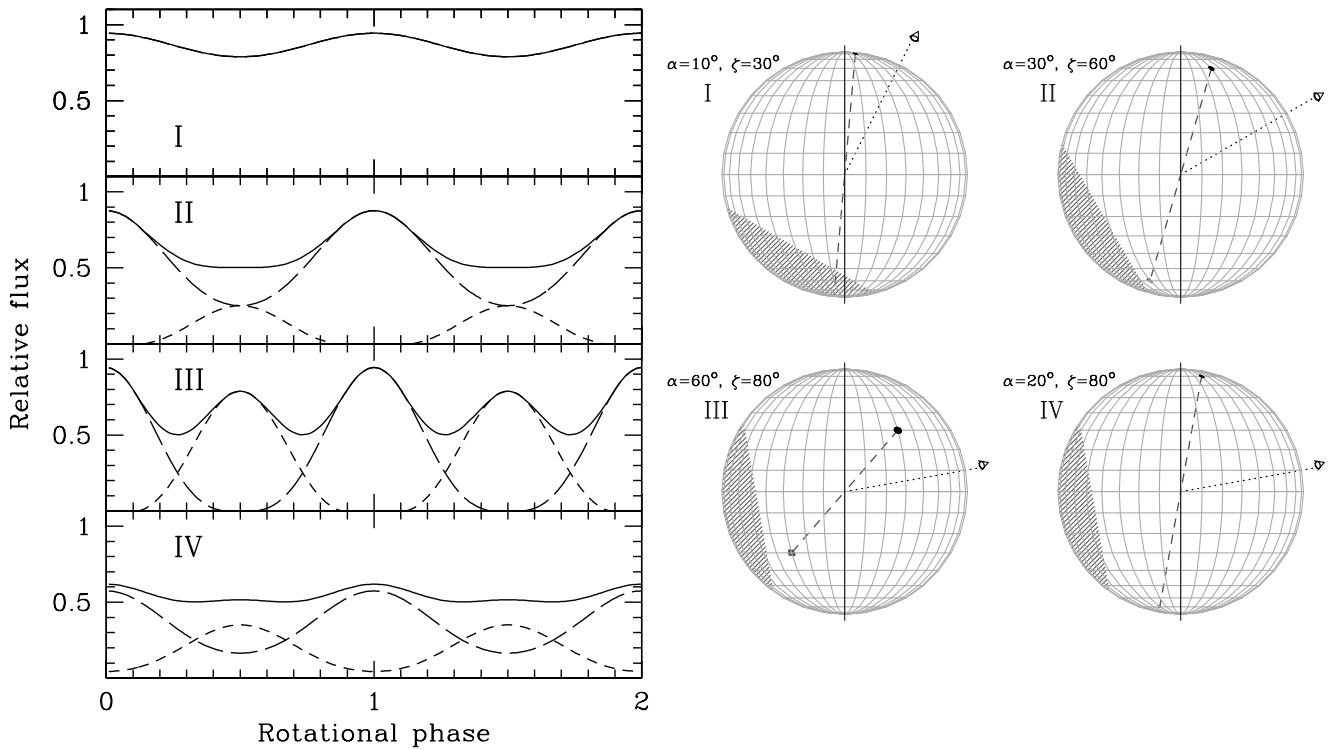}
\caption{(\textit{Left}) Representative synthetic hydrogen atmosphere
  lightcurves for a rotating $M=1.4$ M$_\odot$, $R=10$ km NS with two
  antipodal hot spots for the four lightcurve classes (I-IV, from top
  to bottom, respectively), defined by Beloborodov (2002). The dashed
  lines show the individual flux contribution from the two hot spots
  while the solid line shows the total observed flux.  All fluxes are
  normalized to the value corresponding to a face-on hot spot
  ($\alpha=\zeta=0$). Two rotational cycles are shown for
  clarity. (\textit{Right}) Orthographic map projection of the MSP
  surface for the four pulse profiles. The dashed line shows the
  magnetic axis while the dotted line shows the line of sight to the
  observer. The hatched region corresponds to the portion of the star
  not visible to the observer.}
\end{figure*}

\section{MODEL ANALYSIS}

The specifics of the theoretical model employed in this work are
outlined in \citet{Bog07}. In brief, the model considers a
relativistic rotating compact star with two identical X-ray emitting
hot spots, which in the case of MSPs (presumably) correspond to the
magnetic polar caps. The position of each spot relative to the
observer is given by the angle $\psi$ defined as
\begin{equation}
\cos\psi(t)=\sin \alpha \sin \zeta \cos \phi (t) + \cos \alpha \cos \zeta
\end{equation}
where $\alpha$ is the pulsar obliquity (i.e.~angle between the spin
and magnetic axes), $\zeta$ is the angle between the pulsar spin axis
and the line of sight to the distant observer, and $\phi(t)$ is the
spin phase. We take a non-rotating Schwarzschild metric as a
description of the space-time in the vicinity of the NS and include
Doppler boosting and propagation time delays. This formalism is
remarkably accurate as long as the spin period of the star is
$\gtrsim$3 ms where rotation-induced oblateness is unimportant
\citep{Cad07}. The NS surface is assumed to be covered by an
unmagnetized ($B<10^9$ G), optically-thick H atmosphere
\citep[][]{Zavlin96,McC04}.  The angle-dependent emission pattern of
this atmosphere differs substantially from the isotropic one of a
blackbody, which is particularly evident in the rotation-induced
modulations of the X-ray flux \citep[see Fig.~1 of][]{Bog07}.

%
%
\begin{figure*}[t]
\begin{center}
\includegraphics[width=0.82\textwidth]{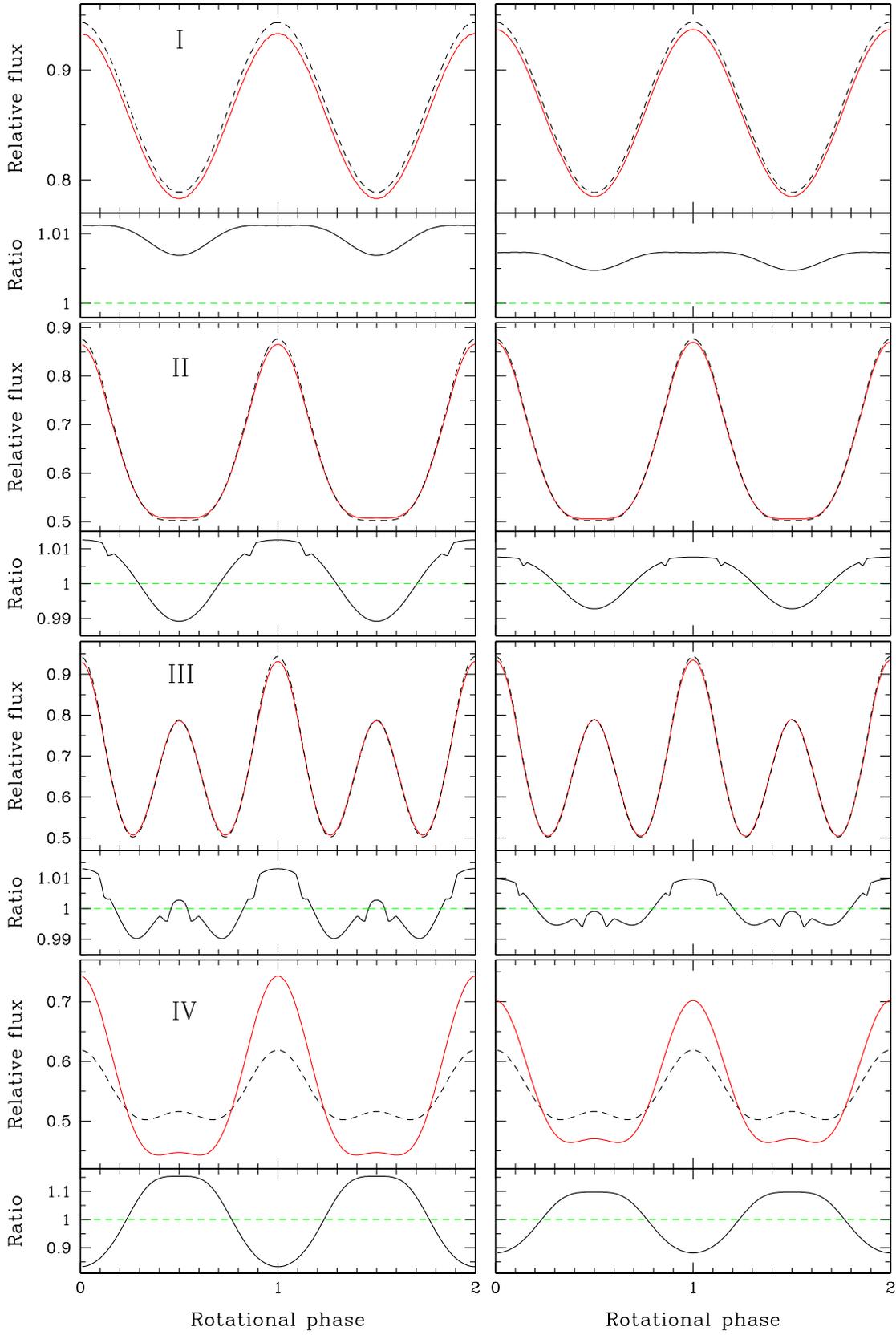}
\end{center}
\caption{Lightcurves for a rotating $M=1.4$ M$_\odot$, $R=10$ km NS
  with two antipodal hot spots with values of $\alpha$ and $\zeta$ as
  in Figure 1.  The \textit{dashed line} shows the idealized case of
  point-like hot spots while the \textit{solid line} is for the case
  of a filled circular hot spot (\textit{left} set of plots) and a
  thin annulus (\textit{right} set of plots), both of radius 2 km and
  $T_{\rm eff}=2\times 10^6$ K.  The bottom panel for each class shows
  the ratio between the two lightcurves shown in the upper panel. All
  fluxes are normalized to the value corresponding to
  $\alpha=\zeta=0$.  Two rotational cycles are shown for clarity.}
\end{figure*}

Using this model we generate both synthetic spectra and pulse profiles
(Fig.~1) for the following parameters: the effective temperatures and
radii of the emission region(s) $T_{1}$, $T_{2}$, $R_1$ and $R_{2}$,
$\alpha$, $\zeta$, and $M/R$. Unless noted otherwise, we fix the mass
to the canonical NS value $M=1.4$ M$_{\odot}$ and vary only $R$.  We
also allow for an off-center magnetic axis by including offsets in the
position of the secondary hot spot in latitude and longitude, $\Delta
\alpha$ and $\Delta \phi$, respectively, from the antipodal position.
The corresponding net offset of the secondary hot spot from the
antipodal position across the NS surface is\footnote{Note that
equation (12) in \citet{Bog07} is valid only if the angle $\alpha$ is
reckoned from the equator towards the spin pole. However, as defined
by convention, the pulsar obliquity $\alpha$ is measured from the spin
pole towards the equator. Thus, the correct expression for the
seondary hot spot offset is given by Equation (2) in this present
paper.}
\begin{equation}
\Delta s = R \cos ^{-1}[\cos\alpha\cos(\alpha+\Delta\alpha)+\sin\alpha\sin(\alpha+\Delta\alpha)\cos\Delta\phi]
\end{equation}
while the total displacement (i.e. impact parameter) of the magnetic
axis from the stellar center is
\begin{equation}
\Delta x = R\sin\left(\frac{\Delta s}{2R}\right)
\end{equation}
We conduct fits to both the spectrum and pulse profile. As pointed out
by \citet{Pavlov97}, this is of importance because the spectral fits
provide tight constraints on $T_{\rm eff}$ and $R_{\rm eff}$ but do
not provide useful information regarding the system geometry or
$M/R$. Conversely, the pulse profiles provide constraints on the
compactness and geometry of the NS but are less sensitive to the
parameters that define the emission spectrum ($T_{\rm eff}$ and
$R_{\rm eff}$).

%
%
\begin{figure}[t]
\begin{center}
\includegraphics[angle=270,width=0.49\textwidth]{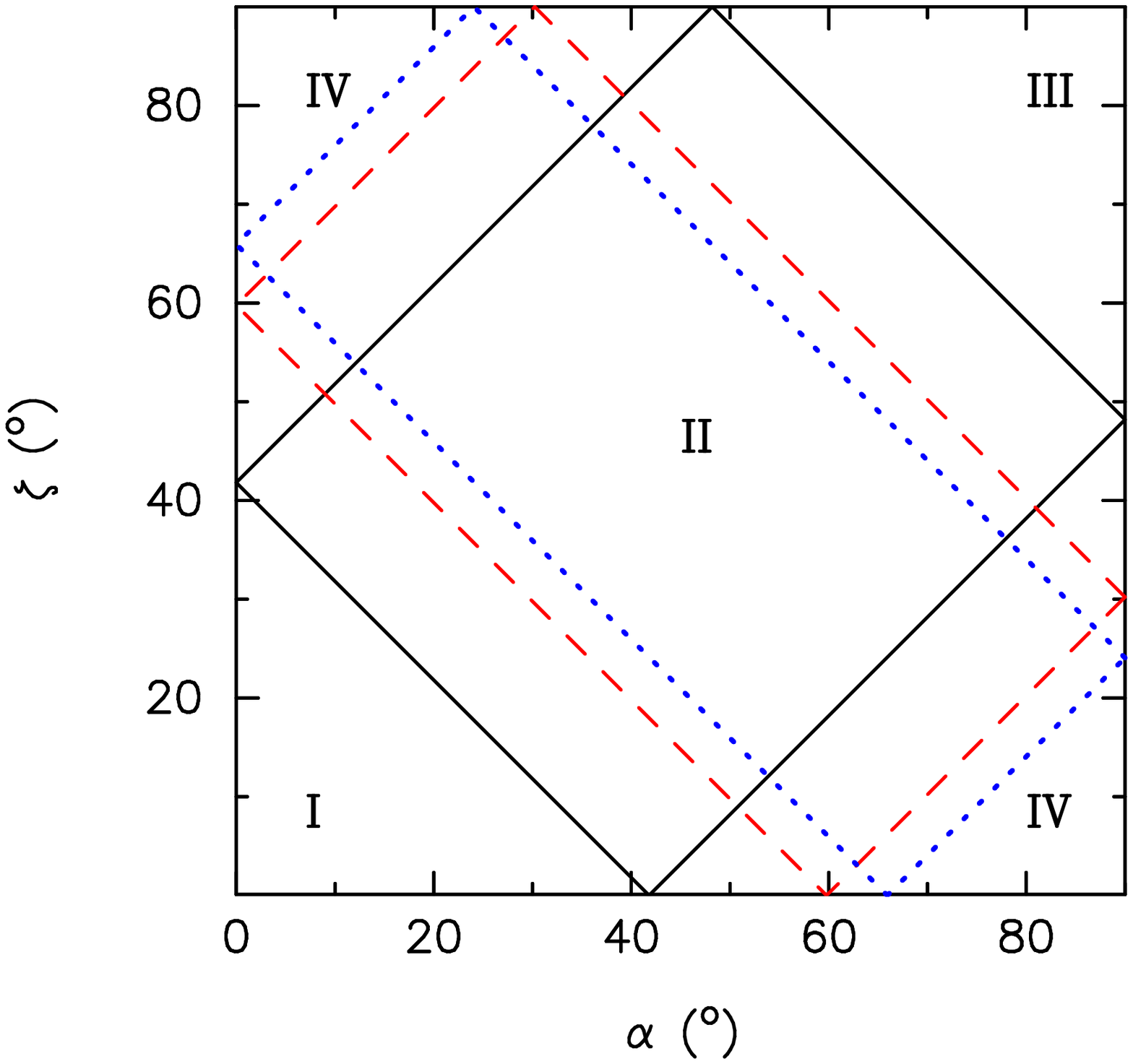}
\caption{The location of pulse profile classes I--IV on the
  $\alpha-\zeta$ plane for MSPs with $R=10$ (\textit{solid}), $12$
  (\textit{dashed}), and $14$ (\textit{dotted}) km assuming $M=1.4$
  M$_{\odot}$. See Figure 1 and text for a definition of each class.}
\end{center}
\end{figure}

\subsection{Extended versus Point-like Emission}

In \citet{Bog07} only point-like hot spots were considered, which was
sufficient for application to the available data for J0437--4715.
Here, we also consider more realistic extended hot spots in order to
ascertain the effect of the uncertain physical size and shape of the
hot polar caps on the observed pulse profiles.  To accomplish this, we
consider a grid of emission spots across the NS surface, which allows
us to construct X-ray emission regions of arbitrary size and shape.
Figure 2 shows the resulting pulse profiles for two extreme cases: 1.~a
uniformly heated, circular polar cap of radius 2 km and 2.~a thin annulus
(with thickness much smaller than its radius) also with a 2 km
radius. The value $R_{pc}=2$ km corresponds to the expected polar cap
radius of a $P=5$ ms pulsar.  Surprisingly, the physical extent and
exact shape of the emission regions do not significantly affect the
observed pulse profile, and only differs from the point-like case by
$\lesssim$1\%, except for class IV where it differs by as much as
$\sim$15\%. The largest differences occur at the turning points of
the lightcurves.

In reality, due to the rotation of the pulsar the polar cap shape is
probably distorted \citep[see, e.g.,][]{Nar83,Biggs90}, with
elongation or compression in the meridional (north-south) direction on
the NS surface. However, even in such instances the differences from a
circular hot spot with the same effective area are relatively small
(few \%).  This also means that for emission regions on the NS surface
comparable in area to that expected for a circular pulsar polar cap,
the X-ray pulse profiles do not provide useful information regarding
the details of the region geometry.  Moreover, the weak dependence of
the pulse profile on the polar cap size implies that the distance to
the pulsar, which is covariant with the emission area through the flux
normalization ($R_{\rm eff}^2/D^2$), does not significantly affect the
pulse profile shape and pulsed fraction. Thus, unlike alternative
methods for measuring NS compactness (e.g., using quiescent low-mass
X-ray binaries), fits to the X-ray pulse profiles of MSPs are not
strongly affected by uncertainties in the distance.  Finally, Figure 2
implies that the pulse profiles are weakly sensitive to broad-band
calibration uncertainties in the detector effective area as
well. Narrow-band deviations from the true effective area of the
instrument (for instance, near absorption edges) are also neglegible
as the pulse profiles cover relatively wide energy intervals (see
\S2.2).

Note that the ambiguity in the exact shape of the polar caps may
ultimately limit the accuracy of the constraint on the desires NS
parameters using the approach described here. Nonetheless, for the
observed spectra of MSPs \citep[see, e.g.,][]{Zavlin06}, the effective
radius of the hotter emission region is found to be of order a few
hundred meters, which compared to the size of the star ($\sim$10 km)
is effectively point-like. Thus, the uncertainty in the true shape of
the emission region can be overcome by considering only the hotter
emission component.

%
%
\begin{figure*}[t]
\begin{center}
\includegraphics[width=0.52\textwidth]{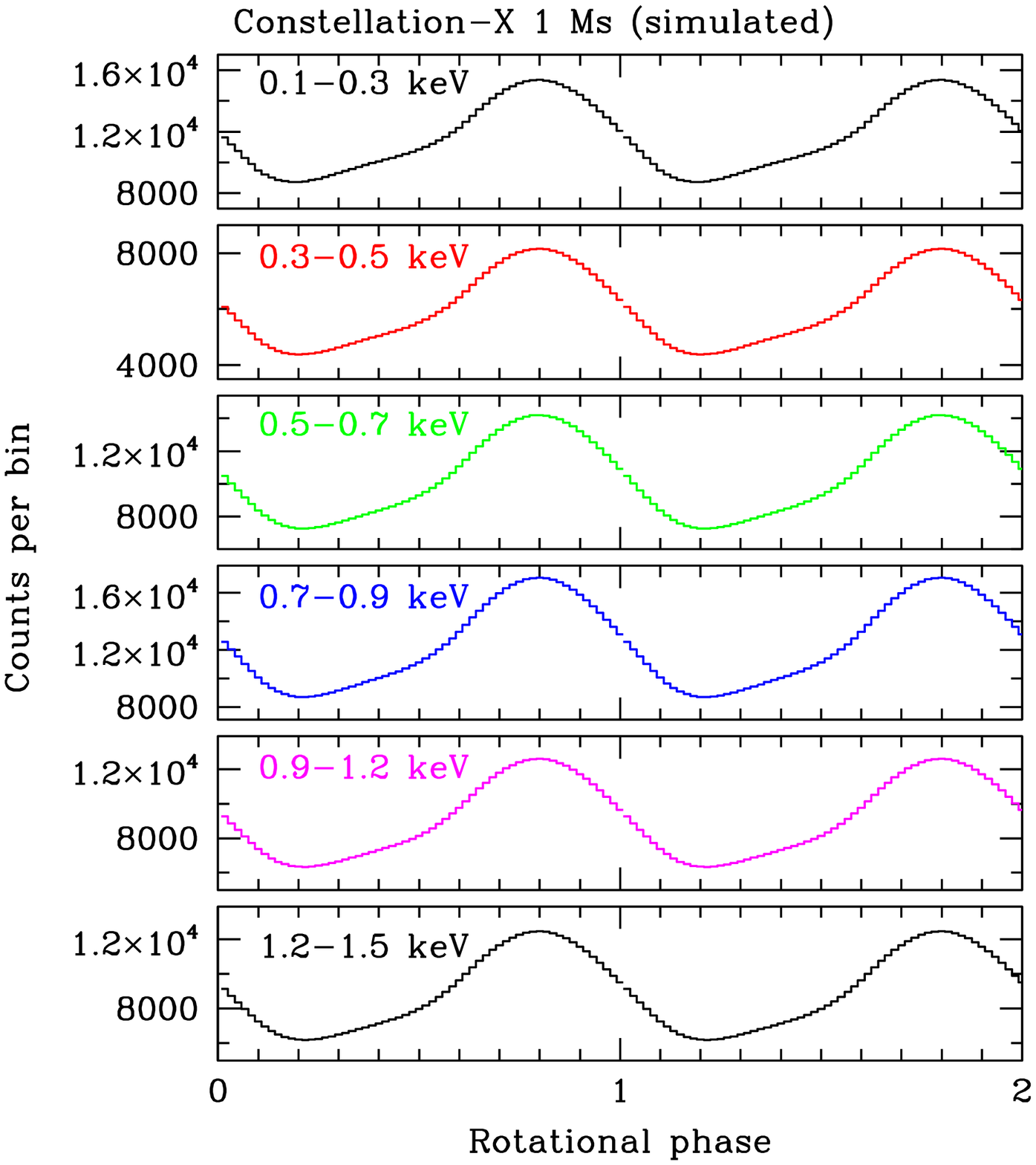}
\includegraphics[width=0.43\textwidth]{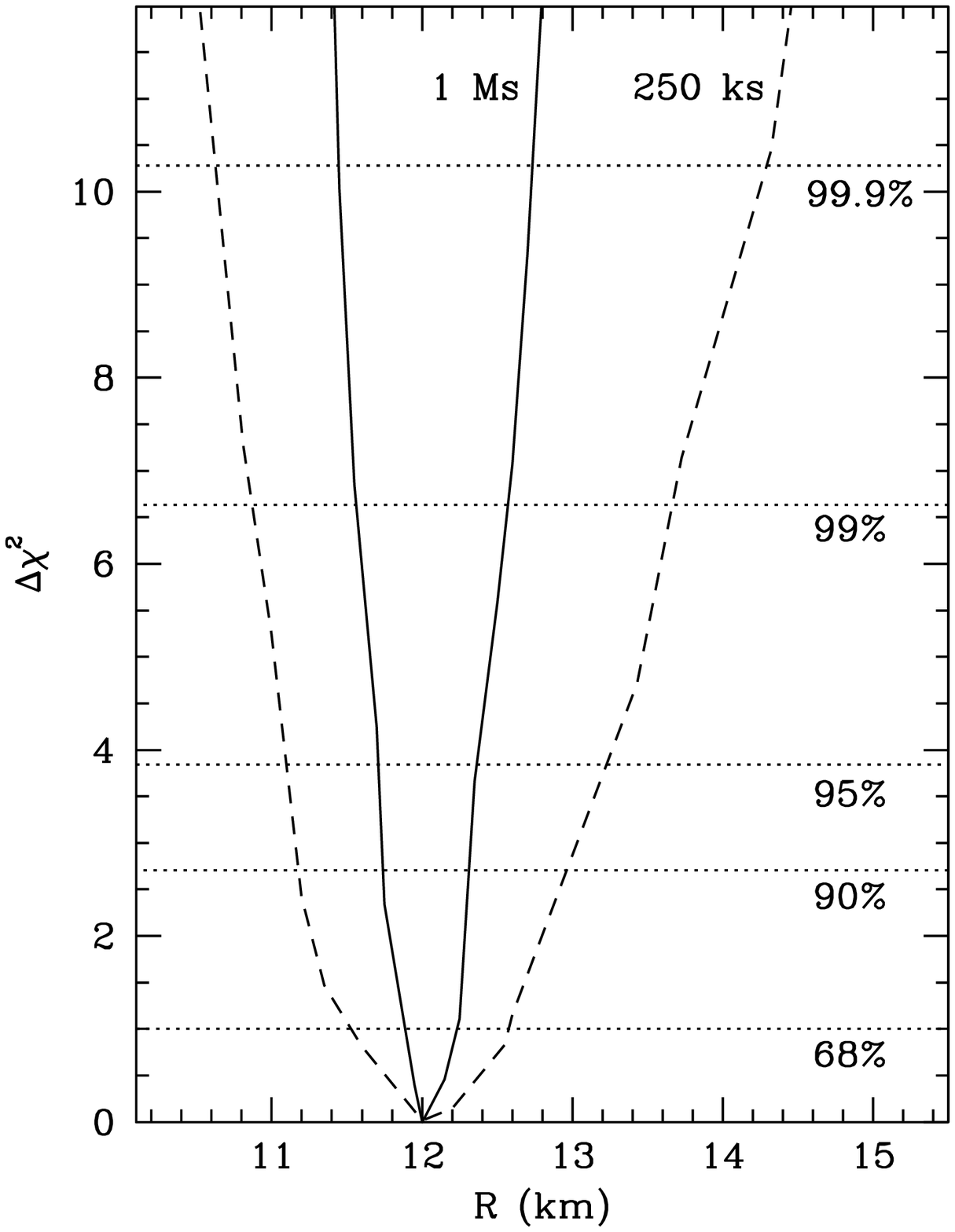}
\caption{(\textit{Left}) Simulated 1-Ms spectroscopic and timing
  observation of PSR J0437--4715 with \textit{Constellation-X} XMS
  assuming $R=12$ km and $M=1.4$ M$_{\odot}$. The choice of phase zero
  is arbitrary. (\textit{Right}) Best fit confidence intervals for $R$
  (assuming $M=1.4$ M$_{\odot}$) from pulse profile fits to simulated
  250-ks and 1-Ms \textit{Constellation-X} XMS observations of PSR
  J0437--4715.}
\end{center}
\end{figure*}

\subsection{Limiting Accuracy of $M/R$ Constraints Using MSPs}

It has been shown that combined X-ray spectroscopic and timing studies
of rotation-powered MSPs can potentially be used to infer $M/R$ of the
NS \citep{Pavlov97,Zavlin98,Bog07}.  Here, we wish to determine
whether with further improvement in data quality, $M/R$ can be
determined to significantly better accuracy.  This is essential if
this method is to be used for reliable measurements of the NS
compactness and ultimately the NS EOS. In addition, it is important to
ascertain how the viewing and magnetic geometries affect the
constraints on the desired NS parameters. For this purpose, we have
carried out a series of fits to simulated X-ray data of MSPs with
greatly improved photon statistics than currently available. A major
advantage of the greatly increased data quality is the possibility of
high signal-to-noise ratio energy-resolved pulse profiles in multiple
energy bands and phase-resolved spectroscopy for several phase
windows.  As an illustrative example, we consider the current version
of the proposed effective area curve of the \textit{Constellation-X}
observatory X-ray Microcalorimeter Spectrometer (XMS)
detector\footnote{See http://constellation.gsfc.nasa.gov/} and assume
a range of exposure times, spanning from 250 ks to 1 Ms.  We generate
representative lightcurves for each of the four distinct types defined
by \citet{Belo02} based on the visibility of the two antipodal hot
spots (see Figs.~1 and 3)\footnote{Note that these are defined for the
antipodal hot spot case, but the regions are similar for offset hot
spots for small displacements.}, assuming both $R=10$ km and $M=1.4$
$M_{\odot}$, representative of strange equations of state, and $R=12$ km
and $M=1.4$ $M_{\odot}$ expected from conventional nucleonic equations of state
\citep{Latt01}.

Class I pulse profiles occur in cases when only the primary hot spot
is observable at all times, while the secondary hot spot is in the
invisible portion of the star for the entire period (see Fig.~1).
Note that for two antipodal polar caps, this can only be achieved for
$R/R_S>1.7$ (or $R>7$ km for $M=1.4$ M$_{\odot}$) since for a more
compact star the entire stellar surface is always visible. The
visibility of only a single hot spot causes large uncertainties
regarding the true geometry and especially the compactness.  This
occurs due to the fact that that visible hot spot is always found at a
small angle with respect to the line of sight where light bending
effects are less pronounced, resulting in weak dependence on $M/R$. As
a result a given class I pulse profile can be reproduced by a fairly
large set of combinations of $\alpha$, $\zeta$, and $M/R$.  We find
that this problem cannot be overcome by deeper observations. Indeed,
for simulated deep exposures ($\sim$1 Ms) with
\textit{Constellation-X} the $90\%$ confidence limits alone encompass
the entire range of plausible NS radii ($7-16$ km for an assumed 1.4
$M_{\odot}$).  Therefore, this class of pulse profiles are not
suitable for tight constraints on $M/R$.

Pulse profiles that occupy region II in Figure 3 exhibit a single
broad pulse per rotation. In this configuration, the primary hot spot
is observable at all times, while the secondary is only visible for a
portion of the spin period. The thermal X-ray pulse profile of PSR
J0437--4715 \citep{Zavlin02,Zavlin06} appears to be in this class.
Figure 4 shows simulated observations of PSR J0437--4715, assuming
$R=12$ km and $M=1.4$ M$_{\odot}$, with \textit{Constellation-X} XMS
and the corresponding best fit confidence intervals for $R$. It is
apparent that with a substantial improvement in photon statistics, $R$
could, in principle, be constrained to better than $10$\%, and
potentially down to $\sim$2--3\% (at 90\% confidence) given sufficient
exposure time. Similar constraints are obtained for $R=10$ km and
$M=1.4$ M$_{\odot}$ as well. Combined with an independent mass
measurement from radio timing observations, this method could lead to
unprecendented constraints on the NS EOS.

Thermal X-ray lightcurves that are found in region III (Fig.~2) are
characterized by two pulse peaks per rotation period. The nearby PSR
J0030+0451 is a good example of this class \citep[][see also
\S3.1]{Beck00,Beck02}.  An advantage of these pulse profiles is that
each of the hot spots provides a dominant contribution to the flux of
one of the two pulses. This feature has several practical
consequences. First, it may, in principle, allow one to determine
whether the two polar caps are identical in terms of size and
temperature. Any significant differences would point to deviations
from a centered dipole, such as displacements of the dipole along the
axial (magnetic north-south) direction or small-scale multipole
contributions. Moreover, the ``sharpness'' of the two pulses is
strongly affected by $M/R$ \citep[see Fig.~3 of][]{Bog07} resulting in
increased sensitivity to the stellar compactness. Thus, class III
appears to be favorable for tight constraints on $M/R$. We find that a
1 Ms simulated \textit{Constellation-X} observation of PSR J0030+0451
permits constraints on $M/R$ down to a $\sim$5\% level (at 90\%
confidence).

Finally, class IV profiles are formed when both hot spots are
observable at all times. As evident in Figure 1, these pulse profiles
tend to have significantly lower fluxes relative to classes I-III
since the hot spots are observed close to edge-on for the entire spin
period.  These pulsars may not be observable at radio wavelengths due
to the large impact angle $|\alpha-\zeta|$, which may exceed the
opening half-angle of the radio emission cone. Such object can only be
detected in X-rays. At present, identifying such objects as MSPs
through X-ray timing is difficult without a known radio counterpart,
although this may be possible with future X-ray observatories (see
\S4.1).  These factors make class IV MSPs the most difficult to study
observationally.  Most importantly, as seen in Figure 2, these
lightcurves are much more sensitive to the size and shape of the hot
spots. This introduces much larger uncertainties (roughly an order of
magnitude greater) than for classes I--III into the spectral and
lightcurve fits, resulting in weaker constraints on $M/R$.

Thus, class II and III pulse profile seem to be most favorable for
tight constraints on $M/R$.  Fortunately, these two classes occupy a
major portion of the $\alpha-\zeta$ plane\footnote{The same holds true
  in the $\cos\alpha-\cos\zeta$ space.} for the plausible range of
$M/R$, implying that most MSPs should fall in these classes.  In
\citet{Bog07} we conducted a detailed study of PSR J0437--4715, a
class II MSP. In \S3, we focus our analysis on two MSPs that exhibit
class III pulse profiles.

\section{Application}

As seen in \S2, for favorable geometries stringent constraints on MSP
parameters are possible.  This suggests that even with fairly limited
photon statistics some useful insight into the MSP properties can be
obtained. Below, we illustrate this through application of our model
to X-ray observations of two nearby MSPs, PSRs J0030+0451 and
J2124--3358. Along with J0437--4715, these are the only thermal MSPs
for which the X-ray pulse profiles are of sufficient quality to allow
meaningful constraints on NS parameters, especially $M/R$. The
spectral analysis for each MSP was performed in
XSPEC\footnote{http://heasarc.gsfc.nasa.gov/docs/xanadu/xspec/} 12.3.0
using the model described above. As shown in
\citet{Pavlov97},\citet{Zavlin98}, and \citet{Bog07}, even the
phase-integrated spectra of MSP are significantly affected by the
choice of $\alpha$ and $\zeta$ due to the energy-dependent
limb-darkening of the atmosphere so it is necessary to consider them
in the spectral fits as well. In order to apply the model to the X-ray
pulse profiles, it was first convolved with the appropriate instrument
response, the encircled energy fraction was taken into account, and
the sky and detector background were added. The fit was performed by
searching the $\chi^2$ hyperspace for the minimum. We fit the pulse
profiles by considering the following parameters: the two temperatures
and effective radii of each hot spot ($T_1$, $T_2$, $R_1$, and $R_2$)
the two angles $\alpha$ and $\zeta$, the stellar radius $R$, the
offsets of the secondary hot spot from the antipodal position ($\Delta
\alpha$ and $\Delta \phi$), and the hydrogen column density along the
line of sight ($N_{\rm H}$).  Unless noted otherwise, in our analysis
we will assume a fixed mass of $M=1.4$ M$_{\odot}$ and allow $R$ to
vary within the range of plausible NS radii for this mass
\citep[$7-15$ km; see, e.g.,][]{Latt01}.
 
%
%
\begin{figure}[t]
\begin{center}
\includegraphics[width=0.49\textwidth]{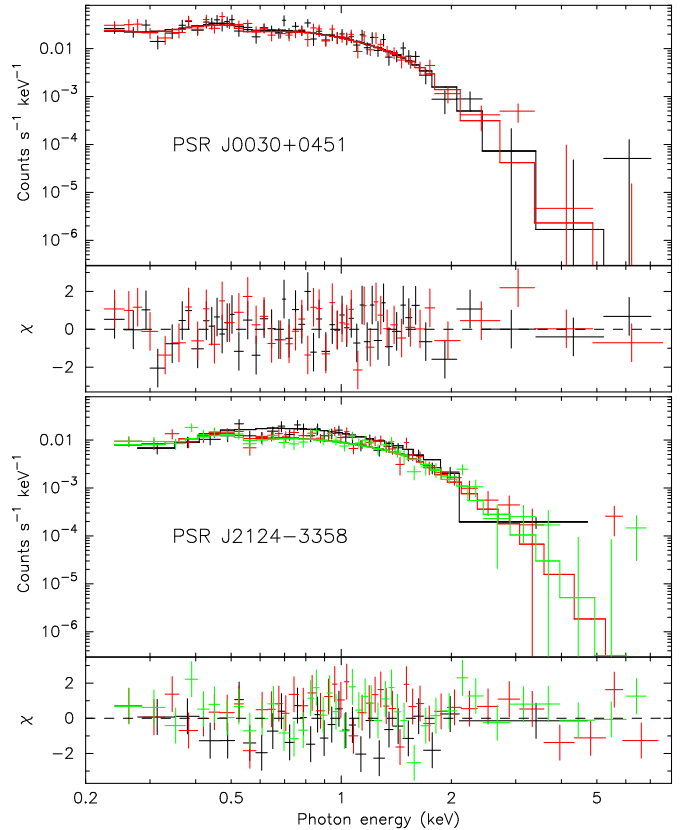}
\caption{(\textit{top}) \textit{XMM-Newton} EPIC-MOS1/2 phase
  integrated spectra of J0030+0451 fitted with a two temperature
  hydrogen atmosphere thermal model. The bottom panel shows the best
  fit residuals. (\textit{bottom}) \textit{Chandra} ACIS-S and
  \textit{XMM-Newton} EPIC-MOS1/2 phase integrated spectra of PSR
  J2124--3358 fitted with a two temperature hydrogen atmosphere
  model.}
\end{center}
\end{figure}

%
%
\begin{figure}[t]
\begin{center}
\includegraphics[width=0.49\textwidth]{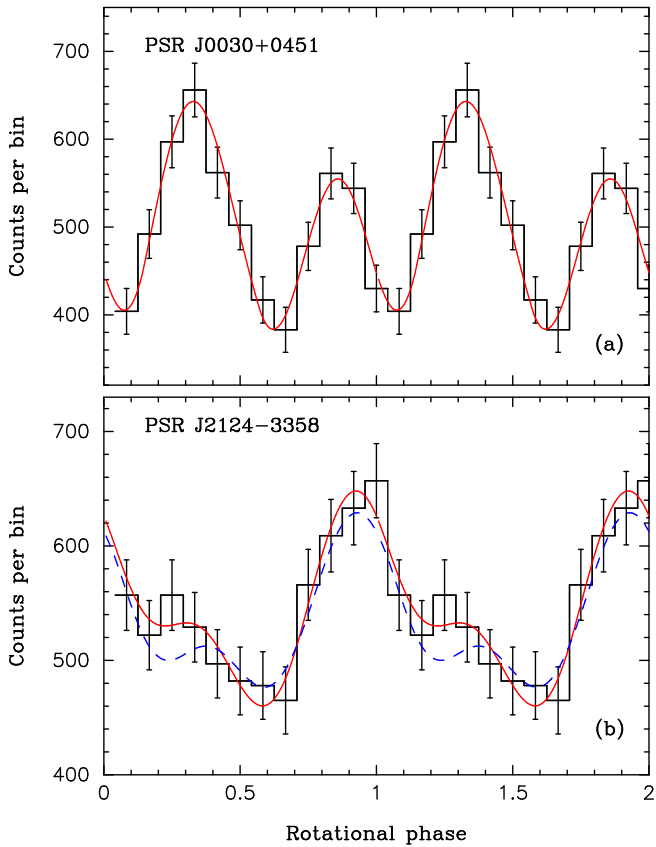}
\caption{(\textit{a}) \textit{XMM-Newton} EPIC-pn pulse profile of
  PSR J0030+0451 for $0.3-2$ keV. The solid line shows the best fit
  model. (\textit{b}) \textit{XMM-Newton} EPIC-pn pulse profile
  of PSR J2124--3358 for $0.3-2$ keV. The solid line shows the best
  fit model, while the dashed line shows the best fit for a centered
  dipole (see text for best fit parameters).}
\end{center}
\end{figure}

\subsection{PSR J0030+0451}

This 4.86-ms solitary pulsar was discovered at radio wavelengths in
the Arecibo drift scan survey \citep{Lom00}. It was subsequently
detected with \textit{ROSAT} during the final days of this mission.
Due to the unstable behavior of the failing PSPC detector, this data
provided no reliable spectral information but revealed a double peaked
pulse profile with pulsed fraction $\sim$50\% \citep{Beck00}.  PSR
J0030+0451 was revisited in 2001 June 19 by \textit{XMM-Newton} for 31
ks \citep{Beck02}. This observation revealed a 0.3--2 keV spectrum
qualitatively similar to that of J0437--4715. In addition, the EPIC-pn
X-ray pulse profile shows a relatively high pulsed fraction
($\sim$50\% for 0.3--2 keV) with two prominent pulses.  \citet{Lom00}
have estimated from radio polarization measurements two possible
pulsar geometries: $\alpha=8^{\circ}$ and $\beta=1^{\circ}$ or
$\alpha=62^{\circ}$ and $\beta=10^{\circ}$, where $\beta$ is the
angular separation of the line of sight with respect to the magnetic
axis at closest approach.  As the former combination of angles cannot
produce the observed double-peaked X-ray profile for all plausible
choices of $M/R$ (see Figs. 1 and 3), even if we allow for substantial
offsets of the secondary hot spots in $\alpha$ and $\phi$,
$\alpha=8^{\circ}$, $\beta=1^{\circ}$ can be definitively ruled
out. This result implies that the observed radio interpulse
\citep{Lom00} most likely originates from the antipodal radio emission
cone, not from the same emission cone as the main pulse.

Figure 5 shows the \textit{XMM-Newton} EPIC-MOS1/2 spectra of PSR
J0030+0451. They are well described by two thermal components. Fitting
a hydrogen atmosphere model to the spectrum yields
$T_{1}=(1.3-2.1)\times 10^6$ K, $R_1=0.01-0.54$ km,
$T_{2}=(0.3-0.7)\times 10^6$ K, $R_2=0.8-3.5$ km, and $F_X=1.5 \times
10^{-13}$ ergs s$^{-1}$ (0.3-2 keV), for assumed $N_{\rm
H}=(1-3)\times10^{20}$ cm$^{-2}$, $M=1.4$ M$_{\odot}$, $R=7-15$ km,
and all combinations of $\alpha$ and $\zeta$ in the range
$0^{\circ}-90^{\circ}$. The ranges quoted are 1$\sigma$ limits. As
expected, the spectral continuum is fit equaly well with a blackbody
model.  \citet{Beck02} have suggested a broken power-law as a possible
alternative interpretation of the spectrum of J0030+0451. However, as
pointed out by \citet{Zavlin07}, this model results in unrealistic
values of $N{_H}$ and is thus unlikely.

The fit to the pulse profile (Fig.~6a) was carried out by fixing
$N_{\rm H}=2\times10^{20}$ cm$^{-2}$ and taking a distance of $D=300$
pc \citep{Lom06}.  For $M=1.4$ M$_{\odot}$, the radius is constrained
to be $R>9.4$ km or, more generally, $R/R_S>2.3$ (68\% confidence) for
all combinations of the other free parameters, with a best fit
$\chi_{\nu}^2=0.81$ (for 3 degrees of freedom). Acceptable fits were
obtained for radii up to $\sim$20 km.  In the case of a blackbody
model, we find that good fits to the X-ray pulse profile require
implausably large stellar radii ($\gtrsim$20 km). As no NS EOS models
predict radii exceeding $\sim$15 km for $M=1.4$ M$_{\odot}$
\citep[see][]{Latt01}, the validity of the blackbody interpretation is
very doubtful. The lower limit on the allowed values of $\alpha$ and
$\zeta$ is found to be $\gtrsim$44$^{circ}$ (68\% confidence),
obtained when the other angle is equal to 90$^{\circ}$. Namely if
$\alpha=90^{\circ}$ then $\zeta=44^{\circ}$ or if $\alpha=44$ then
$zeta=90^{\circ}$. Finally, the currently available data is consistent
with both a centered and a displaced dipole field.

\subsection{PSR J2124--3358}

PSR J2124--3358 is a nearby ($D\approx 250$ pc), isolated pulsar with
$P=4.93$ ms \cite[][]{Bail97} first observed in X-rays with the
\textit{ROSAT} HRI \citep{Beck99}. As the HRI provided no useful
spectral information only a pulse profile was obtained with pulsed
fraction $\sim$33\%. J2124--3358 was later observed with
\textit{Chandra} ACIS-S for 30.2 ks and with \textit{XMM-Newton}
EPIC-MOS1/2 and EPIC-pn for 68.9 and 66.8 ks, respectively
\citep{Zavlin06,Hui06}. As shown by \citet{Zavlin06}, the X-ray
emission from J2124--3358 is also well described by a two-temperature
thermal spectrum. Fitting our hydrogen atmosphere model to the
phase-integrated spectral continuum yields $T_{1}=(1.3-2.4) \times
10^6$ K, $R_1=0.03-0.5$ km, $T_{2}=(0.3-0.8) \times 10^6$ K,
$R_2=0.9-3.1$ km, and $L_X=1.8 \times 10^{30}$ ergs s$^{-1}$ (0.3--2
keV), for assumed $D=250$ pc, $N_H=1\times10^{20}$ cm$^{-2}$, $M=1.4$
M$_{\odot}$, $\alpha=0^{\circ}-90^{\circ}$,
$\zeta=0^{\circ}-90^{\circ}$, and $R=7-15$ km. The uncertainties
quoted represent $\pm$1$\sigma$ ranges.  The derived values are
consistent with the results by \citet{Zavlin06}. Note that the diffuse
X-ray emission detected around this MSP \citep{Hui06} does not
contribute appreciably to the point source MSP emission as its total
luminosity $L_X\sim 1\times 10^{29}$ ergs s$^{-1}$ (0.1--2.4 keV) is
negligible.

The X-ray pulse profile of this MSP (Fig.~6b) exhibits marginal
evidence for a faint secondary peak.  Given that this feature is
evident in both the \textit{ROSAT} PSPC \citep{Beck99} and
\textit{XMM-Newton} EPIC-pn \citep{Zavlin06} pulse profile, it is very
likely genuine.  The fit to the pulse profile (lower panel of Fig.~6b)
was carried out by fixing $N_{\rm H}=1\times10^{20}$ cm$^{-2}$
\citep{Zavlin06} and assuming the dispersion measure derived distance
of $D=250$ pc. Assuming $M=1.4$ M$_{\odot}$, the radius is constrained
to be $R>7.8$ km (68\% confidence), with best fit $\chi_{\nu}^2=$ (for
3 degrees of freedom). The angles $\alpha$ and $\zeta$ are constrained
to be $\ge$12$^{\circ}$ (68\% confidence) for the other angle equal to
$90^{\circ}$ (i.e.~if $\alpha=90^{\circ}$ then $\zeta=12^{\circ}$ or if
$\alpha=12$ then $zeta=90^{\circ}$). Although the suggestive asymmetry
of the pulse profile hints at the presence of an offset dipole, the
poor photon statistics permit a centered dipole configuration. As with
PSR J0030+0451, a blackbody model requires unrealistically large
stellar radii ($\gtrsim$20 km for M=1.4 M$_{\odot}$) for
J2124--3358. Thus, we conclude that a blackbody model does not provide
a valid description of the surface emission properties of PSR
J2124--3358 as well.

\section{DISCUSSION}

%
%
\begin{figure}[t]
\begin{center}
\includegraphics[width=0.45\textwidth]{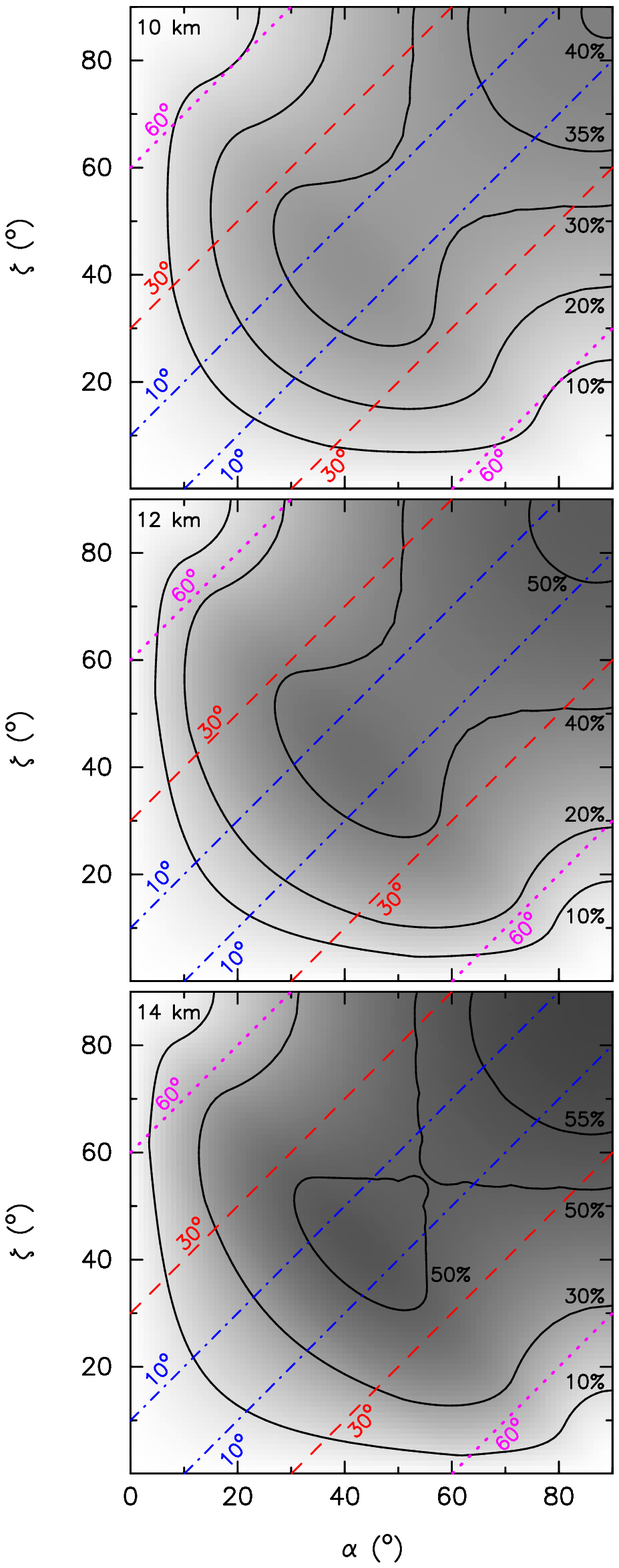}
\caption{Contours of constant pulsed fraction(\textit{solid lines})
  for X-ray H atmosphere emission from two antipodal hot spots in the
  0.3--2 keV band as a function of the pulsar obliquity ($\alpha$) and
  viewing angle ($\zeta$) for a 1.4 M$_{\odot}$ MSP with radius 10,
  12, and 14 km (from \textit{top} to \textit{bottom},
  respectively). The X-ray emission spectrum is taken to be that of
  PSR J0437--4715. The diagonal \textit{dot-dashed}, \textit{dashed}
  and \textit{dotted} lines show $10^{\circ}$, $30^{\circ}$, and
  $60^{\circ}$ opening half-angles of the radio emission cone,
  respectively. All MSPs found between each pair of lines are
  detectable at radio frequencies (see text for details).}
\end{center}
\end{figure}

\subsection{Searches for Radio Quiet MSPs}

The effect of light bending combined with the (nearly) antipodal
configuration of the two MSP hot spots ensure that the thermal
radiation is observable at Earth for any combination of $\alpha$ and
$\zeta$.  In contrast, at radio frequencies a pulsar is not observable
if $|\alpha-\zeta|$ exceeds the opening half-angle $\rho$ of the radio
emission cone.  This brings forth the intriguing prospect of detecting
and identifying such radio quiet MSPs in X-rays using blind pulsation
searches.  With the currently abailable X-ray observatories
(\textit{Chandra} and \textit{XMM-Newton}) this endeavor is rather
difficult due to the intrinsic faintness of MSPs \citep[see,
e.g.,][]{Cam07} and their relatively low X-ray pulsed fractions
($\lesssim$50\%). On the other hand, for the next generation of large
X-ray observatories (\textit{Constellation-X} and \textit{XEUS}), the
proposed the $\gtrsim$10-fold increase in sensitivity makes such a
survey quite feasible.

Figure 7 shows the X-ray pulsed fraction of a 1.4 M$_{\odot}$ MSP with
a 10, 12, and 14 km radius as a function of $\alpha$ and $\zeta$,
assuming a thermal emission spectrum like that of PSR J0437--4715
\citep{Zavlin06,Bog07}.  Also shown are lines delineating the region
of the $\alpha-\zeta$ plane for which a pulsar with a given radio
emission cone width is observable at radio frequencies.  Note that the
true opening angle of the radio cone for a given MSP is not known and
difficult to measure reliably but could be as high as
$\sim$$60^{\circ}$ \citep[see][]{Kra98}.  If we assume a uniform
distribution of pulsar obliquities ($\alpha$) and viewing angles
($\zeta$), for $\rho\lesssim 30^{\circ}$ a substantial portion
($\sim$45\%) of the MSP population is invisible to us in the radio.
On the other hand, if we consider an X-ray timing survey with a
limiting pulsed fraction sensitivity of $\sim$10\%, only $\sim$5--20\%
(depending on $M/R$) of the MSPs will go undetected as pulsed sources
though they will still be detected as X-ray sources (provided they are
not heavily absorbed).  The Galactic population of MSPs may in fact be
preferentially clustered in a certain range of $\alpha$ due to the
poorly understood effects of the accretion and magnetic field
reduction processes during the LMXB phase on the NS \citep[see,
e.g.,][for a review]{Bhatt91}.  A deep X-ray timing survey of nearby
($\lesssim$1--2 kpc) MSPs may, in principle, reveal whether this is
indeed the case.

The detectability of the thermal polar cap emission from MSPs for all
combinations of $\alpha$ and $\zeta$ further suggests that the
\textit{entire} MSP population of certain Galactic globular clusters
\citep[e.g. 47 Tucanae, see][]{Grind02,Bog06a,Cam07} could be detected
using high angular and temporal resolution X-ray imaging and timing
(e.g.~with the aid of the proposed \textit{Generation-X}
observatory). Identifying the whole population of MSPs in a cluster
has important implication for studies of globular cluster evolution
and internal dynamics \citep[see, e.g.,][and references
therein]{Camilo05}.

\subsection{Constraining Magnetic Field Evolution Models}

As shown in \citet{Bog07} and in this present paper, the morphology of
the thermal X-ray pulse profiles of MSPs offer useful insight into the
magnetic field geometry of the pulsar (Fig.~1). This has important
implications for pulsar magnetic field evolution models. For instance,
Ruderman (1991; see also Chen \& Ruderman 1993 and Chen et al.~1998)
has postulated the presence of crustal ``plates'' on the NS surface
formed by shear stresses on the crust caused by neutron superfluid
vortex lines pinned to lattice nuclei (see Fig.~3 of Chen et
al.~1998). The motion of these plates would cause the magnetic fields
of MSPs to migrate across the stellar surface, resulting in either an
aligned magnetic field or one ``pinched'' at the spin pole. However,
these configurations would result in very little ($\lesssim$few
percent) modulation of the thermal X-ray flux due to the close
alignment of the polar caps with the spin axis, regardless of the
viewing geometry. This is at odds with the observed thermal X-ray
pulsed fractions of PSRs J0437--4715, J0030+0451, and J2124--3358,
which are in the range 30--50\%. Thus, although the model of Ruderman
(1991) can reproduce the observed radio properties of MSPs, it is
inconsistent with the observed thermal X-ray pulse profiles of
MSPs. In particular, the X-ray pulse profiles of PSRs J0437--4715,
J0030+0451 and J2124--3358 indicate that the magnetic axes of these
MSPs are significantly misaligned from the spin axis.  This, in turn,
implies that the dipole fields of these (and likely all) MSPs do not
have a tendency to align with the spin axis nor migrate towards one of
the spin poles.

\section{CONCLUSIONS}

We have examined the properties of our model of thermal emission from
hot spots on the surface of a neutron star covered by a hydrogen
atmosphere, relevant for MSPs.  Our investigation has demonstrated
that energy-resolved modeling of the thermal X-ray pulse profiles and
phase-resolved spectroscopy of the continuum emission can, in
principle, be used to determine $M/R$ and the pulsar geometry to high
accuracy, given observational data of sufficient quality and favorable
values of $\alpha$ and $\zeta$. As shown in \S2.1 the thermal pulse
profiles are surprisingly insensitive to the details of the polar cap
size and shape, the distance to the pulsar, and the uncertainty in the
instrument effective area.  This method represents the only feasible
approach of studying MSP magnetic field, surface properties, and
compactness, especially for isolated MSPs.  Barring any deleterious
effect such as additional hidden spectral components \citep[see,
e.g,][and references therein]{Bog06b} this approach can, in principle,
lead to unprecedented insight into NS properties.

Our model is found to be in agreement with the observed emission from
the nearby solitary MSPs J0030+0451, and J2124--3358.  As with PSR
J0437--4715 \citep{Bog07}, the relatively large pulsed fractions
observed in PSRs J0030+0451 and J2124--3358 require the existence of a
light-element atmosphere on the stellar surface and cannot be
reproduced by a blackbody model for realistic NS radii.  For
J0030+0451 and J2124--3358 we are able to place interesting limits on
the allowed compactness of $R>9.4$ km and $R>7.8$ km (68\% confidence)
assuming $M=1.4$ M$_{\odot}$.  Based on our findings in \S2, we expect
deeper observations of this MSP to lead to much tighter constraints on
$M/R$, which in turn may firmly rule out certain families of NS EOS. The
available thermal X-ray pulse profiles also provide useful constraints
on magnetic field models of MSPs. Specifically, the positions of the
magnetic polar caps on the NS surface implied by the X-ray data
indicate that the magnetic field closely resembles the conventional
oblique dipole model of pulsars rather than more exotic field
configurations.

The success of this approach towards elucidating crucial NS properties
motivates further studies of the nearby sample of MSPs with both
\textit{Chandra} and \textit{XMM-Newton}. Moreover, this makes MSPs
particularly important targets for upcoming X-ray mission such as
\textit{Constellation-X} and \textit{XEUS}. The great increase in
throughput of these facilities will allow searches for new MSPs,
detailed observations of a larger sample of known radio MSPs, and
unprecedented constraints on key NS properties, especially the NS EOS.

\acknowledgements We would like to thank Ramesh Narayan, Bryan
Gaensler, Deepto Chakrabarty, Pat Slane, and Alice Harding for
numerous useful suggestions. This work was funded in part by NASA
\textit{Chandra} grant G07-8033A.  The research presented has made use
of the NASA Astrophysics Data System (ADS).


\clearpage


\begin{thebibliography}{}

\bibitem[Bailes et al.(1997)]{Bail97} Bailes, M., Johnston, S., Bell, 
J. F., Lorimer, D. R., Stappers, B. W., Manchester, R. N., Lyne, A. G., 
Nicastro, L., D'Amico, N., \& Gaensler, B. M. 1997, \apj, 481, 386

\bibitem[Becker \& Tr\"umper(1999)]{Beck99} Becker, W. \& Tr\"umper, J. 1999, A\&A, 341, 803 

\bibitem[Becker \& Aschenbach(2002)]{Beck02} Becker, W. \& Aschenbach,
B. 2002, Proceedings of the 270. WE-Heraeus Seminar on Neutron Stars,
Pulsars, and Supernova Remnants, Eds. W. Becker, H. Lech,
J. Tr\"umper, p. 64

\bibitem[Becker et al.(2000)]{Beck00} Becker, W., Tr\"umper, J., Lommen, A. N., \& Backer, D. C. 2000, \apj, 545, 1015

\bibitem[Beloborodov(2002)]{Belo02} Beloborodov, A. M. 2002, \apj, 566, L85

\bibitem[Bhattacharya \& van den Heuvel(1991)]{Bhatt91} Bhattacharya, D. \& van den Heuvel, E. P. J. 1991, Phys. Rep. 203, 1

\bibitem[Biggs(1990)]{Biggs90} Biggs, J. D. 1990, 245, 514

\bibitem[Bogdanov et al.(2006a)]{Bog06a} Bogdanov, S., Grindlay,
J. E., Heinke, C. O., Camilo, F., Freire, P. C. C, \& Becker, W. 2006,
\apj, 646, 1104

\bibitem[Bogdanov et al.(2006b)]{Bog06b} Bogdanov, S., Grindlay,
J. E., \& Rybicki, G. B. 2006, \apj, 648, L55 

\bibitem[Bogdanov et al.(2007)]{Bog07} Bogdanov, S.,
Rybicki, G. B., \& Grindlay, J. E.  2007, \apj, 670, 668

\bibitem[Cadeau et al.(2007)]{Cad07} Cadeau, C., Morsink., S. M.,
Leahy, D., \& Campbell, S. S. 2007, ApJ, 654, 458

\bibitem[Cameron et al.(2007)]{Cam07} Cameron, P. B., Rutledge, R. E.,
Camilo, F ., Bildsten, L., Ransom, S. M., \& Kulkarni, S. R. 2007,
\apj, 660, 587

\bibitem[Camilo \& Rasio(2005)]{Camilo05} Camilo, F. \& Rasio,
  F.~A.~2005, ASP Conf. Ser. Vol. 328: Binary Radio Pulsars,
  ed. F. A. Rasio \& I. H. Stairs (San Francisco: ASP), p. 147

\bibitem[Chen \& Ruderman(1993)]{Chen93} Chen, K. \& Ruderman, M. S. 1993, \apj, 408, 179

\bibitem[Chen et al.(1998)]{Chen98} Chen, K., Ruderman, M., \& Zhu, T. 1998, \apj, 493, 397

\bibitem[Grindlay et al.(2002)]{Grind02} Grindlay, J. E., Camilo, F.,
Heinke, C. O., Edmonds, P. D., Cohn, H., \& Lugger, P. 2002, \apj,
581, 470

\bibitem[Harding \& Muslimov(2002)]{Hard02} Harding, A. K. \&
Muslimov, A. G. 2002, \apj, 568, 862

\bibitem[Hui \& Becker(2006)]{Hui06} Hui, C. Y. \& Becker, W. 2006,
A\&A, 448, L13

\bibitem[Kramer et al.(1998)]{Kra98} Kramer, A., Xilouris, K. M., Lorimer, D. R., Doroshenko, O., Jessner, A., Wielebinski, R., Wolszczan, A., \& Camilo, F. 1998, \apj, 501, 270

\bibitem[Lattimer \& Prakash(2001)]{Latt01} Lattimer, J. M. \&
Prakash, M. 2001, \apj, 550, 426

\bibitem[Lommen et al.(2000)]{Lom00} Lommen, A. N., Zepka, A., Backer,
D. C., McLaughlin, M., Cordes, J. M., Arzoumanian, Z., Xilouris,
K. 2000, \apj, 545, 1007

\bibitem[Lommen et al.(2006)]{Lom06} Lommen, A. N., Kipphorn, R. A.,
Nice, D. J., Splaver, E. M., Stairs, I. H., \& Backer, D. C. 2006,
\apj, 642, 1012


\bibitem[McClintock et al.(2004)]{McC04} McClintock, J. E., Narayan,
R., \& Rybicki, G. B. 2004, \apj, 615, 402

\bibitem[Narayan \& Vivekanand(1983)]{Nar83} Narayan, R. \& Vivekanand, M. 1983, A\&A, 122, 45

\bibitem[Pavlov \& Zavlin(1997)]{Pavlov97} Pavlov, G. G. \& Zavlin,
V. E. 1997, \apj, 490, L91


\bibitem[Ruderman(1991)]{Rud91} Ruderman, M. 1991, \apj, 366, 261

\bibitem[Zavlin et al.(1996)Zavlin, Pavlov, \& Shibanov]{Zavlin96}
Zavlin, V. E., Pavlov, G. G., \& Shibanov, Yu. A. 1996, A\&A, 315, 141

\bibitem[Zavlin \& Pavlov(1998)]{Zavlin98} Zavlin, V. E. \& Pavlov,
G. G. 1998, A\&A, 329, 583

\bibitem[Zavlin et al.(2002)]{Zavlin02} Zavlin, V. E., Pavlov, G. G.,
Sanwal, D.  , Manchester, R. N., Tr\"umper, J., Halpern, J. P., \&
Becker, W. 2002, \apj, 56 9, 894

\bibitem[Zavlin(2006)]{Zavlin06} Zavlin, V. E. 2006, \apj, 638, 951

\bibitem[Zavlin(2007)]{Zavlin07} Zavlin, V.~E. 2007, Ap\&SS, 308, 297

\bibitem[Zhang \& Cheng(2003)]{Zhang03} Zhang, L. \& Cheng, K. S. 2003, A\&A, 398, 639

\end{thebibliography}
\end{document}